\documentclass[12pt]{article}

\usepackage{newtxtext,newtxmath}

\usepackage{graphicx}

\usepackage{hyperref}

\usepackage[letterpaper,margin=1in]{geometry}

\renewenvironment{abstract}
	{\quotation}
	{\endquotation}

\date{}

\makeatletter
\renewcommand{\fnum@figure}{\textbf{\footnotesize Figure \thefigure}}
\renewcommand{\fnum@table}{\textbf{\footnotesize Table \thetable}}
\makeatother

\usepackage{scicite}

\usepackage{url}

\usepackage{multirow}
\usepackage{booktabs}%

\usepackage{xspace}
\newcommand{\idf}[0]{\^{I}le-de-France\xspace}

\def\scititle{
	Do simulated agents move like people?
}
\title{\bfseries \boldmath \scititle}

\author{
	Timothy~LaRock$^{1\ast}$,
	Chen~Zhang$^{1}$,
	J\"urgen~Hackl$^{1}$\and
	\small$^{1}$Department of Civil and Environmental Engineering, Princeton University, Princeton, New Jersey, USA.\and
	\small$^\ast$Corresponding author. Email: larock@princeton.edu\and
}

\begin{document}

\maketitle

\begin{abstract} \bfseries \boldmath
Human mobility is increasingly represented using synthetic populations that offer scalable alternatives when individual-level observations are unavailable or sensitive.
Yet validation typically emphasizes aggregate statistics, which can obscure whether simulated agents traverse transportation networks in ways that resemble real travelers.
Here, we develop a path-centric framework that combines direct path-level comparisons with higher-order network models to compare observed and simulated mobility on a shared metropolitan road network.
Observed and simulated paths share broad statistical regularities and short-range memory.
Beyond these similarities, however, simulated mobility underrepresents long paths, exhibits greater redundancy among long route sequences, covers a smaller and partly different portion of the network, and is more predictable overall.
These discrepancies show that agreement in aggregate mobility patterns does not imply fidelity in how travelers move through the underlying infrastructure.
Higher-order path analysis therefore offers a framework for validating synthetic mobility at the spatial and sequential scales relevant to scientific inference, urban planning, and policy.
\end{abstract}

Human mobility shapes and is itself shaped by society~\cite{Barbosa2018Human, Xu2021Emergence}, facilitating access to opportunity~\cite{Moro2021Mobility, Wang2024Infrequent}, exposure to disease, pollution, and other public health risks~\cite{Kraemer2020effect, Klein2024Characterizing, Wang2026Integrating}, and necessitating large-scale infrastructure management~\cite{Tang2026Human} that itself alters the relationship between humans and other species~\cite{Oliver2026Interacting}.
Yet observing individual movement at the spatial and temporal resolution needed to understand these processes across diverse contexts remains challenging.
Large-scale mobility datasets are expensive to collect, often incomplete or demographically biased, and inherently sensitive because detailed trajectories can reveal identities, habits, and private information~\cite{Yabe2024Enhancing,Armenante2025Protecting,Chen2026biases}. 
These limitations motivate the growing use of agent-based simulations and synthetic populations as substitutes for directly observed mobility~\cite{Eubank2004Modelling, Nagel2009Agentbased,Chakirov2014Enriched,Horni2016MultiAgent,Hackl2019Epidemic, Schlosser2021Finding, Horl2021Synthetic, Nguyen2021overview, Metz2024Interactive}. 
By combining population-level statistics with behavioral and transportation models, such simulations promise fine-grained mobility data without requiring the collection or release of individual mobility trajectories.

However, the scientific value of synthetic mobility data depends on whether simulated agents move like real people. 
Most simulation studies evaluate realism through aggregate quantities such as trip counts, travel times, mode shares, origin\textendash{}destination flows, or distance distributions~\cite{Horl2021Synthetic}.
Agreement in these quantities is important, but it does not establish that a simulation reproduces how individuals navigate through the underlying transportation network, nor does it necessarily capture the inherently multi-scale patterns observed in human mobility~\cite{Balcan2009Multiscale, Alessandretti2020scales}. 
Two datasets can have similar trip-length distributions or aggregated traffic volumes, even while individual agents travel fundamentally different paths through the network.
Aggregate validation may therefore conceal behavioral discrepancies that propagate into downstream analyses of accessibility, exposure, congestion, resilience, or epidemic transmission. 
This raises a central question: \emph{when, and in what respects, can an agent-based mobility simulation serve as a faithful surrogate for human mobility?}

\begin{figure}[!p]
	\centering
	\includegraphics[width=\textwidth]{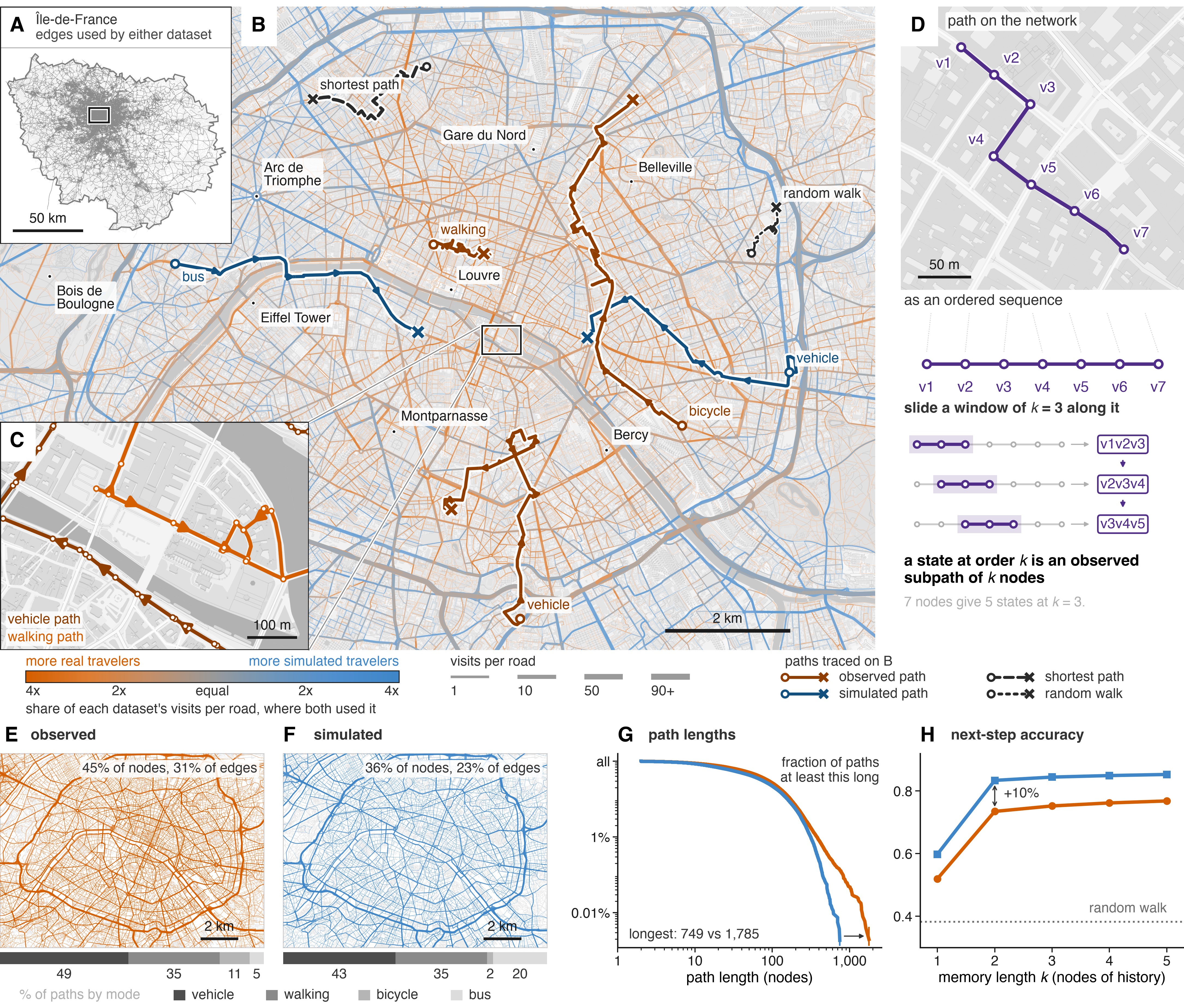}
\caption{\footnotesize\textbf{Path-based comparison of observed and simulated mobility on a shared metropolitan road network.}
\textbf{(A)} Road network of the \idf{} region, with edges used by either dataset highlighted; the box marks the central Paris region shown in (B).
\textbf{(B)} Road-network use in central Paris, combining visitation frequency and relative use by observed and simulated mobility, with representative paths for the transportation modes considered in the analysis, alongside shortest-path and random-walk baselines derived from the observed data.
\textbf{(C)} Detailed view of observed vehicle and walking paths, with network nodes shown along each path.
\textbf{(D)} Higher-order representation of a network path.
Sliding a window of \(k\) consecutive nodes along the ordered node sequence defines the states of an order-\(k\) model, so that each state corresponds to an observed subpath of \(k\) nodes.
\textbf{(E, F)} Network use by observed and simulated mobility, respectively, over the same region and spatial scale. 
At this level of aggregation, the two show only minor differences.
Annotations report the fractions of network nodes and edges visited, and the bars show transportation-mode composition.
\textbf{(G)} Complementary cumulative distributions of path length, measured in nodes, showing the longer upper tail of observed mobility.
\textbf{(H)} Next-step prediction accuracy as a function of memory length \(k\), showing the gain from path history and the greater overall predictability of simulated mobility.
Basemaps use CARTO Positron with data from OpenStreetMap contributors.}
	\label{fig:overview}
\end{figure}

We address this question by shifting the unit of analysis from locations and flows to the paths that individuals take through a transportation network (Fig.~\ref{fig:overview}). 
A path records not only which streets or intersections are visited, but also the order in which they are traversed.
This distinction matters because human movement is purposeful and history dependent~\cite{Manley2015Shortest,Zhu2015People, Lima2016Understanding}.
A traveler approaching an intersection is not choosing the next road solely based on their current location; the choice is also influenced by their preceding route from the origin point, their destination, and the mode in which they are traveling (\emph{e.g.}, private car or public bus).
Conventional network models neglect this structure by implicitly treating movement as a memoryless process in which the next step depends only on the outgoing links of the present node~\cite{Scholtes2017When,Lambiotte2019networks}.
A simulation can therefore reliably reproduce the visitation frequencies of individual road segments while failing to reproduce the sequential logic through which those road segments are combined into journeys.

To expose such differences, we develop a path-centric framework based on higher-order network science~\cite{Lambiotte2019networks}. 
Higher-order network models encode sequences of visited nodes as states in an expanded network, allowing memory-dependent movement to be represented and analyzed using standard network methods~\cite{Scholtes2017When, Lambiotte2019networks}.
In this representation, model ``order'' corresponds to the length of path history retained: an order-\(k\) model distinguishes visits to the same physical location according to the preceding \(k-1\) locations.
By explicitly encoding this memory, higher-order models preserve sequential dependencies that are lost when mobility is reduced to isolated edges, spatial cells, points of interest, or origin\textendash{}destination counts.
Moreover, the appropriate order can be estimated statistically from the data rather than specified a priori, while the resulting higher-order networks provide natural generative models of path data that can be used for statistical testing~\cite{Scholtes2017When,Gote2023Predicting,LaRock2020HYPA}.

This perspective leads to a more demanding test of synthetic mobility.
Instead of asking only whether real participants and simulated agents use the same physical network, we ask whether they traverse the network in a similar way, explore comparable portions of the network, and exhibit similar levels of memory-dependence and route predictability.
The resulting framework identifies discrepancies that remain invisible to conventional aggregate comparisons and localizes them at the level of nodes, edges, and paths.

We apply this framework to observed and simulated mobility in the \idf region of France.
The 2025 NetMob Data Challenge provides GPS trajectories from a large mobility survey designed to represent travel within the region~\cite{Chasse2025NetMob25}.
We map-match these trajectories to the road network to obtain sequences of traversed network locations.
We compare them with paths generated using Open \idf, a detailed synthetic population and agent-based transportation model implemented in MATSim~\cite{Horl2021Synthetic}.
Because the observed and simulated movements are represented on the same shared physical network, differences between them can be evaluated directly rather than being confounded by incompatible spatial resolutions or network representations.

Our comparison reveals a consequential divide between statistical similarity and behavioral realism.
Observed and simulated trajectories share broad mobility regularities and exhibit comparable short-range memory, indicating that the agent-based model captures important structure beyond the topology of the road network alone.
Yet this agreement breaks down when the diversity and predictability of complete paths are considered.
Simulated agents underrepresent rare long journeys, explore a more restricted portion of the transportation network, visit nodes and edges with different frequencies, and follow routes that are substantially more predictable than those of observed travelers.
The simulation therefore represents routine and strongly directed mobility more faithfully than the heterogeneous and atypical movements present in real behavior.

These findings have implications beyond the specific simulation and region studied here.
Synthetic populations have enormous potential where direct mobility observations are unavailable, inaccessible, or too sensitive to distribute.
However, if synthetic trajectories systematically suppress rare, exploratory, or weakly constrained behavior, analyses based on them may underestimate the diversity of human behavior and the range of possible system states.
This limitation may be especially important for applications governed by uncommon journeys or deviations from routine, for example in the case of disaster response~\cite{Tang2026Human}.
Validation must therefore evaluate not only whether a model reproduces average behavior, but also whether it preserves the behavioral diversity and sequential structure relevant to the intended application in a meaningful way.

Our approach complements several established traditions in mobility research.
Spatial interaction and mobility models characterize movements between locations without necessarily resolving the infrastructure paths connecting them~\cite{Erlander1990gravity,Simini2012universal,Louail2015Uncovering,Pappalardo2015Returners,Feng2018DeepMove,Simini2021Deep,Schlapfer2021universal,Bontorin2025Mixing}. 
Studies of street networks analyze the topology, morphology, and structural importance of urban infrastructure, often independently of the detailed trajectories unfolding on those networks~\cite{Lee2017Morphology,Kirkley2018betweenness,Boeing2025Modeling,Boeing2025Topological}. 
Other approaches aggregate mobility into spatial grids or visits to points of interest~\cite{Krumme2013predictability,Jiang2012Clustering,Hasan2013Spatiotemporal,Toole2015path,Zeng2017Visualizing,Chen2020Using,Fu2025Linking,Mauro2026urban}. 
These representations are valuable for many questions, but aggregation severs the connection to underlying infrastructure, makes distinct paths indistinguishable, and erases the sequential dependencies that determine how people navigate physical infrastructure.

Here, we couple the topology of the transportation network with the higher-order dynamics of movement through it. 
To our knowledge, this provides the first head-to-head higher-order comparison of observed and agent-based mobility paths represented on a shared metropolitan road network. 
More broadly, the framework establishes a transferable test of synthetic mobility that distinguishes reproducing aggregate patterns from reproducing human behavior. 
Our results show that these forms of agreement are not equivalent and that apparent realism at the population-level can mask systematic simplification at the level of individual paths.

\subsection*{Results}
We compare two datasets of human mobility paths in the \idf{} region: 50,661 observed paths from the NetMob 2025 Data Challenge~\cite{Chasse2025NetMob25} and 53,716 simulated paths generated by the Open \idf{} MATSim model~\cite{Horl2021Synthetic}.
In both datasets, mobility is represented as sequences of nodes traversed on the same OpenStreetMap (OSM) road network~\cite{OpenStreetMapcontributors2017Planet}, allowing observed and simulated paths to be compared directly on a common network.
We focus on surface transportation modes and therefore exclude train, tram, and subway travel; path counts by mode are reported in Table~\ref{tab:mode-breakdown}, and the full data-processing pipeline is described in Methods.

For comparison, we construct two baseline path datasets that capture contrasting forms of network traversal: \textit{random walks} and \textit{shortest paths}.
Random walks represent uninformed movement determined only by local network topology, whereas shortest paths represent globally directed movement between fixed origins and destinations.
Together, these baselines provide contrasting benchmarks for assessing the structure of observed and simulated mobility.

We organize the comparison across four complementary dimensions of mobility paths.
First, we compare path-length distributions in terms of network nodes and physical distance, showing that simulated mobility underrepresents long paths but exhibits greater physical distance per node among shorter paths.
Second, we quantify complementary dimensions of subpath diversity and find that observed mobility contains a broader range of long route sequences with less repetition, while subpath occurrence frequencies are similarly concentrated in both observed and simulated mobility.
Third, we compare network coverage and behavioral alignment, revealing that simulated agents explore a smaller portion of the transportation network and increasingly diverge from observed travelers as longer subpaths are considered.
Finally, we use higher-order network models to quantify sequential structure and route predictability, showing that both datasets exhibit short-range memory but that simulated paths are systematically more predictable than observed human mobility.

\begin{figure}[!ht]
  \centering
  \includegraphics[width=\textwidth]{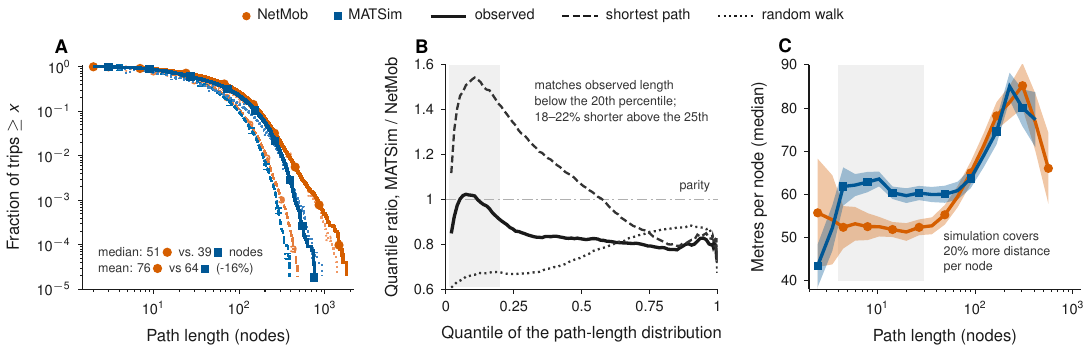}
\caption{\footnotesize\textbf{Simulated mobility underrepresents long paths, but traverses more physical distance per node among shorter paths.}
\textbf{(A)} Complementary cumulative distribution functions of path length, measured in nodes, for observed and simulated mobility and their corresponding shortest-path and random-walk baselines.
Each random walk starts at the same node and contains the same number of nodes as its paired mobility path before filtering, whereas each shortest path connects the same origin and destination.
\textbf{(B)} Ratio of simulated to observed physical-distance quantiles for the mobility, shortest-path, and random-walk distributions.
Values below parity indicate shorter physical distances in simulated mobility at the corresponding quantile.
The shaded region highlights the lowest 20\% of the distribution, where observed and simulated physical-distance quantiles are closely aligned; above approximately the 25th percentile, simulated paths are consistently shorter.
\textbf{(C)} Median physical distance per node within logarithmically spaced bins of path length, with approximate 95\% confidence intervals for the median.
The shaded region highlights paths containing 5 to 20 nodes, for which simulated mobility traverses systematically more physical distance per node than observed mobility.
}
  \label{fig:lengths}
\end{figure}

\subsection*{Simulated mobility underrepresents long paths}

We first compare the path-length distributions of observed and simulated mobility and their corresponding baselines, considering both the number of traversed network nodes and physical distance (Fig.~\ref{fig:lengths}).
Observed and simulated paths both exhibit heavy-tailed node-length distributions, but the simulated distribution is shifted toward shorter paths (Fig.~\ref{fig:lengths}A).
The median path contains 51 nodes in the observations and 39 in the simulation, while the corresponding means are 76 and 64 nodes, respectively.
However, the difference in node count does not translate uniformly into physical distance.
The physical-distance distributions are closely aligned below approximately the 20th percentile, whereas simulated paths are about 18--22\% shorter across much of the distribution above the 25th percentile (Fig.~\ref{fig:lengths}B).
At the same time, among paths containing 5 to 20 nodes, simulated paths traverse approximately 17--20\% more physical distance per node than observed paths (Fig.~\ref{fig:lengths}C).
The difference is particularly pronounced in the upper tail of the node-length distribution (Fig.~\ref{fig:lengths}A).
The longest observed path traverses 1{,}833 nodes and spans 161\,km, compared with 753 nodes and 114\,km for the longest simulated path.
Moreover, 93 observed paths traverse more nodes than any path generated by the simulation.
The difference extends beyond these most extreme observations: the 99th-percentile path length is 403 nodes for observed mobility compared with 315 nodes for simulated mobility.
Thus, while the simulation reproduces the broad heavy-tailed form of the observed path-length distribution, it systematically underrepresents long paths.
The shortest-path baselines are likewise shifted toward shorter node sequences, whereas the random-walk distributions remain close to their paired mobility distributions because path length is matched by construction before network simplification (see Methods).
We consider the potential contributions of route generation, sampling, and data construction to these differences in the Discussion.

\begin{figure}[!ht]
  \centering
  \includegraphics[width=\textwidth]{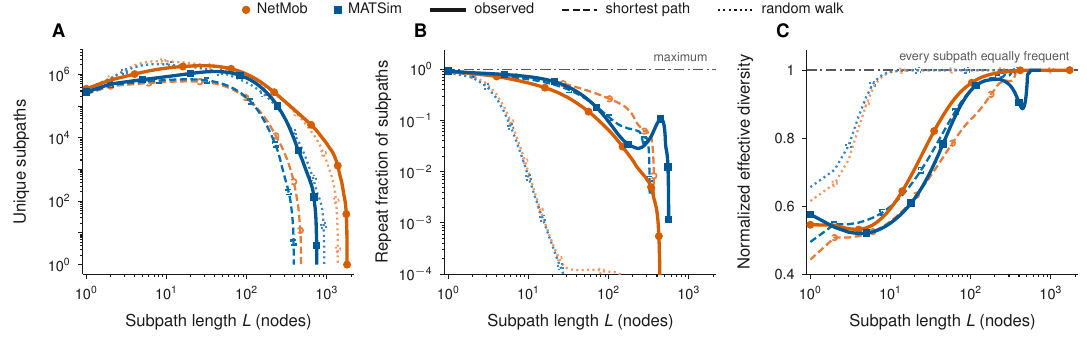}
\caption{\footnotesize\textbf{Observed mobility exhibits greater diversity among long route sequences.}
\textbf{(A)} Number of unique subpaths as a function of subpath length \(L\) for observed and simulated mobility and their corresponding shortest-path and random-walk baselines.
\textbf{(B)} Repeat fraction, defined as one minus the ratio of unique subpaths to total subpath occurrences at each \(L\).
Values near 1 indicate extensive repetition, whereas values near 0 indicate that most subpath occurrences are unique.
\textbf{(C)} Normalized effective diversity of the subpath occurrence-frequency distribution, measuring how evenly occurrences are distributed among the unique subpaths observed at each \(L\).
A value of 1 indicates that all distinct observed subpaths occur equally often, whereas smaller values indicate that occurrences are increasingly concentrated among a subset of these subpaths.}
  \label{fig:diversity}
\end{figure}

\subsection*{Observed mobility exhibits greater subpath diversity}
Paths through the same network can partially overlap by traversing limited sequences of common edges while differing substantially in their overall structure.
Mobility path datasets may visit nodes and edges at similar rates but in very different sequences, resulting in different levels of subpath diversity.
We therefore quantify \emph{unique subpaths} to compare the diversity of route sequences across observed and simulated mobility and their corresponding baselines (Fig.~\ref{fig:diversity}A).
The number of unique subpaths at length \(L\) is directly related to the path-length distribution, since only paths of length \(L' \geq L\) can contribute subpaths of length \(L\).
The number of unique subpaths also reflects the regularity of the paths themselves, since longer paths with greater overlap produce fewer unique subpaths than paths of similar length that share little or no substructure.

Unique-subpath counts peak at \(L=27\) for observed mobility and \(L=39\) for simulated mobility (Fig.~\ref{fig:diversity}A).
For both corresponding random-walk baselines, the peak occurs at length 7, whereas the shortest-path baselines peak at length 17.
Beyond these peaks, unique subpath counts decline rapidly for the shortest-path baselines, while the random-walk baselines follow a pattern more similar to the observed mobility data.
However, because these curves reflect both the path-length distribution and repeated route structure, the raw counts alone cannot isolate differences in subpath diversity.

To separate repeated route structure from differences in path length, we compare the number of unique subpaths with a combinatorial upper bound determined solely by the full path-length distribution.
For a given subpath length \(L\), this bound represents the maximum number of unique subpaths that could be generated by the observed paths of length \(L' \geq L\), assuming no repetition within or overlap between paths (see Methods for details).
We express this comparison as a repeat fraction, defined as one minus the ratio between the observed number of unique subpaths and this upper bound (Fig.~\ref{fig:diversity}B).
A repeat fraction near zero indicates that nearly all subpaths permitted by the combinatorial bound are unique, whereas larger values indicate greater repetition or overlap.

The random-walk baselines rapidly approach repeat fractions near zero, indicating comparatively little repeated subpath structure, whereas the shortest-path baselines retain substantially greater repetition over longer sequences (Fig.~\ref{fig:diversity}B).
This contrast is expected because shorter shortest paths can recur as subpaths of longer shortest paths.
Observed and simulated mobility show broadly similar reductions in repetition across short and intermediate subpath lengths, but their behavior separates in the tail.
At longer supported subpath lengths, the repeat fraction continues to decrease for observed mobility but rises again for simulated mobility, indicating greater redundancy among long simulated route sequences (Fig.~\ref{fig:diversity}B).

Figure~\ref{fig:diversity}C complements these richness- and repetition-based comparisons by measuring normalized effective diversity, which indicates how evenly subpath occurrences are distributed among the distinct route sequences that are observed (see Methods).
The closer normalized effective diversity is to 1, the more uniformly the visitations are spread across all observed subpaths, while values below 1 indicate concentration of visitations among a subset of subpaths.
At short subpath lengths, normalized effective diversity is approximately 0.5--0.6 for both observed and simulated mobility, indicating that occurrences are concentrated among a subset of the available unique subpaths.
The measure increases toward 1 as \(L\) grows, while the random-walk baselines approach this limit much more rapidly, consistent with a more even distribution of occurrences across their unique subpaths.
Observed and simulated mobility follow broadly similar trends over most of the supported range, despite the clearer differences in unique-subpath counts and repeat fraction shown in Figs.~\ref{fig:diversity} A and B.
Together, these results indicate that the principal difference in subpath diversity arises from the number and repetition of distinct long route sequences rather than from a broad systematic difference in how evenly those sequences are used.

\begin{figure}[!ht]
  \centering
  \includegraphics[width=\textwidth]{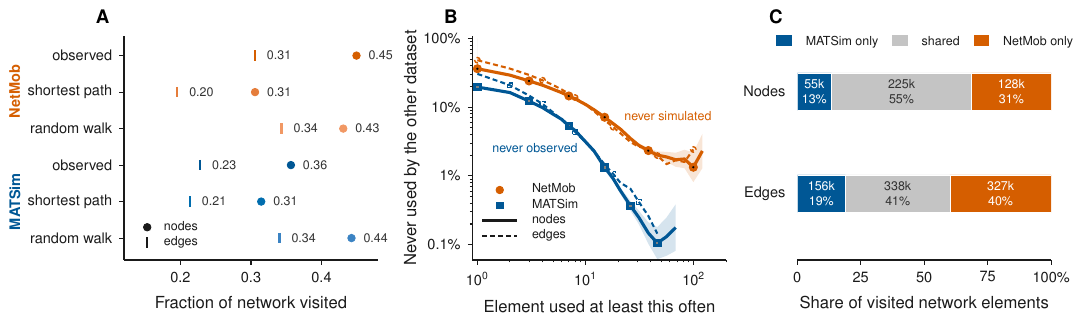}
\caption{\footnotesize\textbf{Observed mobility covers more of the road network, while substantial fractions of visited elements differ between datasets.}
\textbf{(A)} Fractions of nodes and edges in the simplified road network visited by observed and simulated mobility and their corresponding shortest-path and random-walk baselines.
\textbf{(B)} For network elements visited at least the indicated number of times in one mobility dataset, the fraction that is never visited in the other.
Node-level estimates are shown with 95\% Wilson confidence intervals.
\textbf{(C)} Composition of the union of visited nodes and edges, partitioned into elements used only by simulated mobility, shared by both datasets, or used only by observed mobility.
Counts are reported in thousands together with their percentage of the corresponding union.}
  \label{fig:coverage}
\end{figure}

\subsection*{Behavioral alignment weakens across scales}
We next compare the extent and composition of road-network use in observed and simulated mobility before examining alignment across longer route sequences.

Network coverage is quantified as the fraction of nodes and edges in the simplified road network visited by each mobility process (Fig.~\ref{fig:coverage}A).
Observed mobility covers approximately 45\% of network nodes and 31\% of edges, compared with 36\% of nodes and 23\% of edges in the simulation.
Observed mobility therefore covers a larger portion of the network despite containing fewer total paths.
Together with the differences in path length and subpath diversity, this reduced coverage shows that simulated mobility uses a narrower portion of the road network.

Coverage differences also persist when network elements are stratified by how frequently they are used.
As the minimum visitation threshold increases, the fraction of elements completely absent from the other dataset decreases, indicating greater agreement for heavily used network elements (Fig.~\ref{fig:coverage}B).
Across most of the supported range, however, this unmatched fraction is larger for elements used by observed mobility but absent from simulated mobility than for the reverse comparison.
Within the union of visited network elements, 55\% of nodes and 41\% of edges are shared between observed and simulated mobility.
Of the remaining elements, 31\% of nodes and 40\% of edges are used only by observed mobility, compared with 13\% of nodes and 19\% of edges used only by simulated mobility (Fig.~\ref{fig:coverage}C).
Thus, simulated mobility not only covers less of the network overall but is also absent from a substantial subset of the network elements used by observed mobility.

\begin{figure}[!ht]
  \centering
  \includegraphics[width=\textwidth]{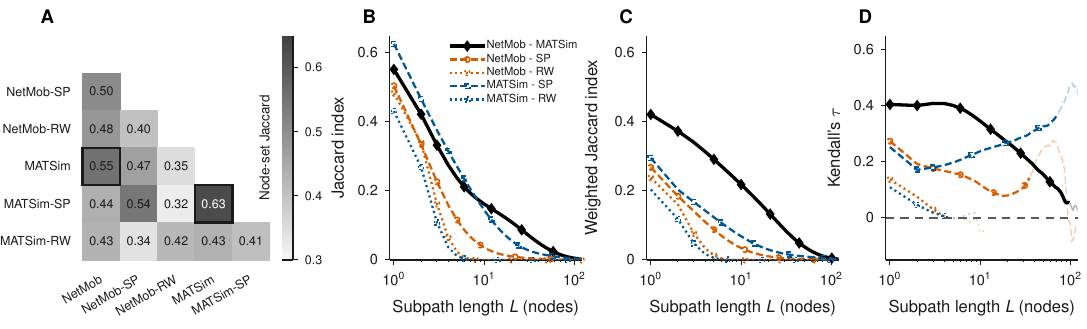}
\caption{\footnotesize\textbf{Observed and simulated mobility align at short scales but diverge across longer route sequences.}
\textbf{(A)} Pairwise Jaccard similarity between the sets of nodes visited by observed and simulated mobility and their corresponding shortest-path and random-walk baselines.
\textbf{(B)} Jaccard similarity between sets of unique subpaths of length \(L\).
\textbf{(C)} Weighted Jaccard similarity accounting for subpath occurrence frequencies.
\textbf{(D)} Kendall's \(\tau\) rank correlation between subpath occurrence frequencies.
Estimates are shown faded where fewer than \(10^4\) subpaths are shared or the correlation is not significant at \(p < 0.05\).}
  \label{fig:alignment}
\end{figure}

%
We next extend the comparison from overall network coverage to overlap in the specific nodes and ordered subpaths used by each dataset (Fig.~\ref{fig:alignment}).
This matters directly for our central question: high overlap would support the use of simulated mobility as a surrogate for observed network usage, whereas low overlap would indicate that the two datasets rely on different parts of the infrastructure.
At the node level, observed and simulated mobility have a Jaccard similarity of 0.55, indicating substantial but incomplete overlap in the locations they visit (Fig.~\ref{fig:alignment}A).
The largest pairwise node overlap occurs between simulated mobility and its shortest-path baseline, with a Jaccard similarity of 0.63.
Extending the comparison to ordered subpaths, the Jaccard similarity between observed and simulated mobility declines steadily as subpath length \(L\) increases and approaches zero for the longest sequences considered (Fig.~\ref{fig:alignment}B).
Across much of the intermediate range, observed and simulated mobility remain more similar to each other than to their corresponding random-walk baselines and, over much of the range, their shortest-path baselines.

Weighting the comparison by subpath occurrence frequency yields the same qualitative pattern: observed and simulated mobility remain the most closely aligned pair across much of the supported range, but their similarity declines toward zero as \(L\) increases (Fig.~\ref{fig:alignment}C).
While weighted Jaccard similarity captures differences in visitation frequency, it does not assess whether the relative ranking of subpaths is preserved across datasets.
We therefore compare the frequency rankings of shared subpaths using Kendall's \(\tau\) as \(L\) increases (Fig.~\ref{fig:alignment}D).
Estimates are shown faded where fewer than \(10^4\) subpaths are shared or the correlation is not significant at \(p < 0.05\).
The rank correlation between observed and simulated mobility is approximately \(0.4\) at short subpath lengths, remains relatively stable over the first few values of \(L\), and then declines toward zero as longer, well-supported subpaths are considered.
Rank correlations with the corresponding random-walk baselines are lower and decay toward zero over shorter scales; beyond \(L = 4\) they are statistically indistinguishable from zero even though tens of thousands of subpaths are still shared.

A different pattern emerges for the shortest-path baselines.
For simulated mobility, rank correlation with the shortest-path baseline increases with \(L\) and eventually exceeds the correlation between observed and simulated mobility.
For observed mobility, the correlation with its shortest-path baseline decreases across the well-supported range; the apparent increase at the largest \(L\) occurs only after the number of shared subpaths becomes small and should therefore be interpreted cautiously.
Together, the overlap and rank-based comparisons show that observed and simulated mobility align most strongly at short scales and progressively diverge across longer route sequences, while the frequency ordering of longer simulated subpaths increasingly resembles that induced by shortest-path routing.

\begin{figure}[!ht]
  \centering
  \includegraphics[width=\textwidth]{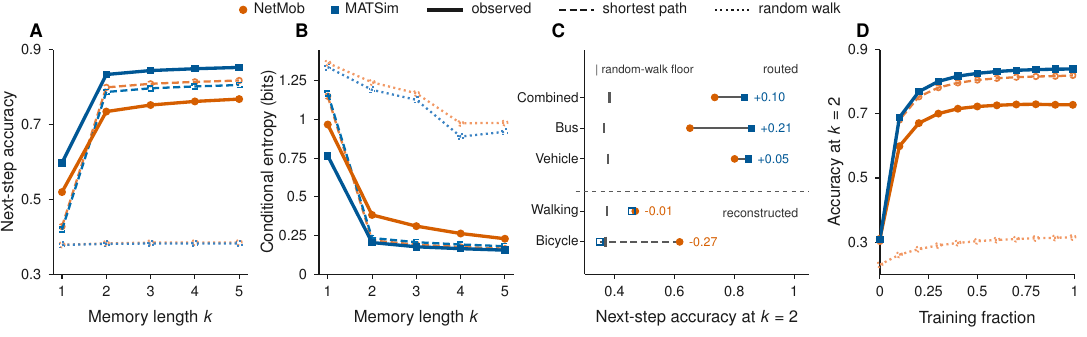}
\caption{\footnotesize\textbf{Simulated mobility is more predictable overall, with mode-specific reversals.}
\textbf{(A)} Next-step prediction accuracy for the combined transportation modes as a function of model memory length \(k\), averaged over ten stratified train--test splits.
\textbf{(B)} Conditional entropy of the empirical next-step distributions for the combined transportation modes as a function of memory length; lower values indicate a more constrained continuation given the preceding path history.
\textbf{(C)} Observed and simulated next-step prediction accuracy at \(k=2\) for each transportation mode.
Labels report the simulation--observation difference in accuracy, with the corresponding observation-derived random-walk baseline shown for reference.
Vehicle paths are explicitly routed in the simulation and bus paths follow scheduled transit routes, whereas bicycle and walking paths are reconstructed as shortest paths between their recorded origins and destinations.
\textbf{(D)} Combined-mode prediction accuracy at \(k=2\) as a function of the fraction of paths used for training; the rightmost point uses 99\% of paths for training.}
  \label{fig:predictability}
\end{figure}

\subsection*{Simulated mobility is more predictable}
We finally examine the \emph{predictability} of mobility paths using higher-order network models.
We formulate a next-step prediction task that uses models with increasing memory length to predict the next node of a partially observed path (see Methods).
Higher prediction accuracy indicates stronger sequential regularity, because the next movement is more strongly determined by the preceding path history.

For the combined transportation modes, next-step prediction accuracy increases sharply from memory length \(k=1\) to \(k=2\) for both observed and simulated mobility and changes only modestly at higher orders (Fig.~\ref{fig:predictability}A).
A memory length of two is also identified as optimal by the likelihood-based order estimation~\cite{Scholtes2017When}, while increasing the memory length further provides only marginal gains in prediction accuracy at the cost of more model parameters.
Despite this similar dependence on memory length, combined-mode accuracy at \(k=2\) is approximately 10 percentage points higher for simulated than observed mobility (Fig.~\ref{fig:predictability}C).

The corresponding shortest-path baselines show a similar increase with memory length, with consistently higher prediction accuracy for the observation-derived baseline than for the simulation-derived baseline.
As expected for a memoryless process, prediction accuracy for the random-walk baselines remains substantially lower and changes little with increasing memory length (Fig.~\ref{fig:predictability}A).

The same difference appears in the conditional entropy of the next-step distributions (Fig.~\ref{fig:predictability}B).
Conditional entropy measures the uncertainty of the next movement given the preceding path history, with higher values indicating greater ambiguity among possible next steps and lower values indicating a more constrained continuation.
For both datasets, conditional entropy decreases sharply when the memory length increases from one to two and continues to decline more gradually at higher orders, showing that additional path history reduces uncertainty about the next step.
For the combined transportation modes, observed paths exhibit higher conditional entropy than simulated paths at every memory length considered, indicating greater uncertainty in their next-step distributions.
Thus, both measures indicate greater aggregate predictability in simulated mobility.

Mode-specific accuracy at \(k=2\) shows that the overall predictability gap is not uniform across transportation modes (Fig.~\ref{fig:predictability}C).
The simulation exceeds the observations by 0.21 in prediction accuracy for bus paths and by 0.05 for vehicle paths, whereas walking differs by only \(-0.01\) and bicycle paths show a pronounced reversal of \(-0.27\).
Notably, vehicle and bus paths show higher predictability in the simulation, whereas walking and bicycle paths reconstructed using shortest paths do not.
For bus travel, the greater simulated predictability may partly reflect fixed route structure, whereas observed trips labeled as bus can contain segments traveled by other modes because multi-mode trips are classified according to their main mode (see Methods).
For walking and bicycle travel, simulated paths are reconstructed as shortest paths rather than explicitly routed; because the predictor is destination-blind, such shortest-path reconstruction does not necessarily imply high next-step predictability from a partial path.
Data sparsity may also contribute to the bicycle result, since the simulated dataset contains only about 1.2k bicycle paths compared with 5.7k observed paths.
%
%
%
These mode-specific differences suggest that comparisons involving non-vehicle mobility require particular care, an issue we return to in the Discussion.

Finally, we test whether the combined-mode predictability difference can be explained by the amount of training data by fixing the memory length at \(k=2\) and varying the training fraction (Fig.~\ref{fig:predictability}D).
With no training data, prediction accuracy is identical for the two datasets.
Prediction accuracy increases rapidly with additional training data for both observed and simulated mobility, but the two curves plateau without converging.
A clear gap remains even when 99\% of the available paths are used for training, indicating that limited training data alone does not explain the greater predictability of the simulation.

Taken together, these results show that both datasets contain higher-order sequential structure, but simulated mobility is more predictable overall, with the magnitude and direction of this difference varying across transportation modes and route-generation procedures.

\section*{Discussion}
Our results provide a qualified answer to whether agent-based mobility simulations can serve as faithful surrogates for observed human mobility.
At a broad level, observed and simulated paths exhibit similar statistical and sequential regularities: both follow heavy-tailed path-length distributions, are best represented by a higher-order model with memory length \(k=2\), and differ clearly from memoryless random walks.
Yet this agreement weakens when mobility is examined at the level of the paths travelers take through the transportation network.
The simulation underrepresents the longest journeys, exhibits greater redundancy among long subpaths, covers a smaller and partly different portion of the network, and produces mobility that is more predictable overall, despite mode-specific reversals.
Together, these findings show that agreement in aggregate mobility patterns does not ensure fidelity in how travelers move through the underlying infrastructure.

One important source of these discrepancies may be the difficulty of representing heterogeneous and individually specific mobility behavior in large-scale simulations.
Synthetic populations encode the demographic, activity, and behavioral information available from the data and models used to construct them, from which agents generate travel according to prescribed decision rules and constraints~\cite{Horl2021Synthetic}.
Such models may reproduce regular and frequently observed mobility patterns while still underrepresenting less common combinations of activities, destinations, and route choices.
This interpretation is consistent with our results, as the largest discrepancies emerge for longer paths and subpaths, for which observed mobility contains more distinct route sequences and less repeated structure than simulated mobility, even though the concentration of subpath frequencies among the subpaths is broadly similar in both datasets.
Importantly, such movements should not simply be interpreted as noise.
Trips to less common destinations and deviations from population-level travel patterns can reflect purposeful and even routine behavior for an individual, despite being difficult to represent statistically at the population-level~\cite{Napoli2026One}.
The challenge for synthetic mobility is therefore not simply to introduce additional randomness, but to capture the heterogeneity in activities, constraints, and preferences that shapes individually meaningful movement patterns, even when these are uncommon at the population-level.
This distinction may have practical consequences, as infrequent mobility behavior has been associated with socioeconomic outcomes in cities~\cite{Wang2024Infrequent}.

Differences in network usage are particularly important when synthetic mobility is intended to support infrastructure-level inference.
A simulation may reproduce trip counts, travel distances, or other population-level statistics while directing travelers through systematically different parts of the transportation network.
Our analysis shows that this discrepancy extends beyond overall network coverage, with observed and simulated paths differing in the nodes and edges they use, how frequently they use them, and which longer route sequences are most commonly traversed.
The appropriate level of validation depends on the intended use of the synthetic data.
For applications concerned primarily with aggregate mobility demand, path-level discrepancies may be of limited consequence.
For analyses of congestion, accessibility, exposure, resilience, or other processes that depend directly on \textit{where} and \textit{how} people traverse infrastructure, however, faithfully reproducing network usage becomes essential.
Synthetic mobility should therefore be validated at the spatial and sequential scales relevant to the intended application rather than against aggregate statistics alone.

Not all observed discrepancies should be attributed to the behavioral model itself, as some may arise from differences in sampling and data construction.
The NetMob and MATSim datasets represent the underlying population in fundamentally different ways, and neither exhaustively samples mobility in the region~\cite{Chasse2025NetMob25,Horl2021Synthetic}.
Finite sampling is particularly important for heavy-tailed mobility, because including or excluding a comparatively small number of highly mobile individuals or long journeys can substantially affect the observed path distribution and network coverage.
In principle, simulated agents could be calibrated more directly using observed participant data or larger mobility datasets.
Such calibration, however, complicates validation because using the same detailed path data for both model tuning and evaluation risks circularity.
More fundamentally, relying on detailed observed paths for calibration undermines one of the principal advantages of synthetic mobility, namely its potential use where individual-level observations are unavailable or cannot be distributed.
Recent work on learning agent-based models directly from data offers a promising route toward richer behavioral models~\cite{Monti2023learning}.
Such approaches will nevertheless require evaluation on held-out behaviors and path-level statistics to determine whether they generalize beyond the observations used for calibration.

Our results also demonstrate why higher-order network representations provide a useful complement to conventional mobility validation.
A first-order network records which nodes and edges are traversed but not the sequence in which they are combined into paths.
By retaining path history, higher-order models expose sequential regularities that remain invisible in first-order representations.
Both observed and simulated mobility exhibit clear memory effects, as a memory length of two is favored by model selection, prediction accuracy improves substantially with additional path history, and conditional entropy declines as longer histories are considered.
At the aggregate level, these measures reveal stronger sequential predictability in simulated mobility, with higher next-step accuracy and lower conditional entropy, although prediction differences vary across transportation modes.
The comparison with random walk and shortest path baselines further helps distinguish sequential structure associated with mobility behavior from patterns imposed by the topology of the road network alone.
In particular, as subpath length increases across the well-supported range, the growing rank correlation between simulated paths and their shortest-path baseline shows that the frequency ordering of simulated route sequences becomes progressively more similar to that of shortest paths.
Higher-order analysis therefore provides not only an additional validation metric, but also a way to identify the scales at which observed and simulated mobility begin to differ.

Several methodological aspects of the analysis warrant caution.
Most importantly, walking and cycling paths in the MATSim dataset are not generated as explicit network routes during the simulation.
Because these modes are implemented through teleportation between origins and destinations, we reconstruct their paths using shortest path routing (see Methods).
Their sequential structure therefore partly reflects this reconstruction procedure rather than the underlying MATSim behavioral model.
This limitation is especially relevant to the mode-specific prediction results.
It also highlights the need for mobility simulations that represent pedestrian and cyclist route choice explicitly, including preferences related to safety, infrastructure quality, convenience, and leisure travel~\cite{Hunter2021Effect,Wang2024CycleTrajectory, Zafri2026Advancing}.
The observed data introduce a different source of uncertainty.
Multi-mode NetMob trips are classified according to their main mode, so a path labeled as walking, cycling, or bus travel may contain segments traveled using another mode.
In addition, the observed paths are derived by map matching noisy GPS measurements, and 3{,}692 trajectories for which no reliable match could be obtained were excluded from the analysis.
These limitations are particularly relevant for pedestrian movement, which is not always constrained to the mapped road network.

The predictability analysis deliberately conditions on path history but not on the traveler's destination.
This allows us to isolate sequential regularity, but human mobility is intrinsically goal-directed, and knowledge of the destination may substantially constrain the set of plausible next steps.
Destination-aware prediction could therefore help separate predictability attributable to path history from that attributable to knowledge of the destination.
Likewise, the random walk and shortest path baselines used here were designed as simple behavioral reference points while preserving selected properties of the original paths.
Future work could consider richer null models, including weighted, biased, or higher-order random walks, to isolate more specific mechanisms underlying path structure.
Finally, our analysis considers a single metropolitan region, a single simulation framework and configuration, and only road-based transportation modes.
Extending the framework across cities, simulation configurations, transportation systems, and additional behavioral attributes will be necessary to determine which of the patterns identified here generalize across mobility systems.

More broadly, our findings argue for evaluating synthetic mobility at the level at which it will ultimately be used.
Agent-based simulations can reproduce important statistical regularities and short-range sequential structure in human mobility, but agreement at these levels does not guarantee that simulated agents reproduce the diversity, network usage, and predictability of observed paths.
Higher-order path analysis provides a general framework for identifying such discrepancies by coupling the topology of the transportation network with the sequential structure of movement through it.
As privacy constraints and limited data availability drive greater reliance on synthetic populations, establishing when simulated mobility is a credible surrogate will require validating not only where and how far people travel, but also how they move through the networks that connect those locations.

\begin{table}[!ht]\footnotesize\centering
	\begin{tabular}{llccc}
	\toprule
		Mode & NetMob Mode Name~\cite{Chasse2025NetMob25} &  NetMob Count & MATSim & Combined Total \\
		\midrule
		\multirow[t]{6}{*}{Vehicle}
		& PRIV\_CAR\_DRIVER & 20,915 \\
		& PRIV\_CAR\_PASSENGER & 2,952 \\
		& LIGHT\_COMM\_VEHICLE & 44 \\
		& TAXI & 195 \\
		& TWO\_WHEELER & 530 \\
		\cmidrule(lr){2-5}
		& Total Vehicle & 24,636 & 22,861 & 47,497 \\
		\midrule
		\multirow[t]{4}{*}{Bicycle}
		& BIKE & 3,693 \\
		& ELECT\_BIKE & 1,685 \\
		& ELECT\_SCOOTER & 348 \\
		\cmidrule(lr){2-5}
		& Total Bicycle & 5,726 & 1,223 & 6,949 \\
		\midrule
		Walking & WALKING & 17,571  & 18,685 & 36,256 \\
		\midrule
		Public Transportation & BUS & 2,728 & 10,947 &  13,675 \\
		\midrule
		Total & \multicolumn{1}{c}{\textendash{}} & 50,661 & 53,716 & 104,377 \\
	\bottomrule
	\end{tabular}
	\caption{\footnotesize Paths per dataset broken down by transportation mode, after the network-simplification step. The Combined Total column is the sum of the NetMob and MATSim counts.}
	\label{tab:mode-breakdown}
\end{table}

\section*{Materials and Methods}

\subsection*{Study design}
We compare observed and simulated human mobility at the level of paths traversed through a common metropolitan road network.
Observed mobility is derived from GPS measurements collected as part of the NetMob 2025 Data Challenge~\cite{Chasse2025NetMob25}, whereas simulated mobility is generated using the Open \idf{} implementation of the Multi-Agent Transport Simulation (MATSim) framework~\cite{Horl2021Synthetic}.
Both datasets are projected onto the same OpenStreetMap (OSM) road network and processed using the same network-simplification procedure before analysis.
We restrict the comparison to surface transportation modes represented on the road network, including vehicle, bus, bicycle, and walking trips, and exclude train, tram, and subway travel.
The resulting datasets contain 50{,}661 observed and 53{,}716 simulated paths after preprocessing.
We compare the datasets along four complementary dimensions: path length, subpath diversity, network coverage and path alignment, and sequential predictability.
Throughout these analyses, memoryless random-walk and shortest-path baselines provide reference cases for local memoryless traversal and directed origin--destination routing, respectively.
Higher-order network models are used to quantify sequential dependencies through optimal-order estimation, next-step prediction, and conditional entropy.

\subsection*{Observed mobility data}
The NetMob 2025 Data Challenge dataset contains mobility observations from 3{,}320 participants living in the \idf{} region~\cite{Chasse2025NetMob25}.
Participants carried dedicated GPS devices and recorded the purpose and transportation mode of trips taken during a seven-day observation period between October 2022 and May 2023.
The dataset contains more than 80{,}000 trips together with demographic information and trip-level metadata.
Individual movements are represented as temporally ordered GPS coordinates that were preprocessed by the data provider to address outliers, missing observations, trip segmentation, and participant privacy~\cite{Chasse2025NetMob25}.
Many trips contain multiple transportation modes.
For the mode-specific analyses, we assign each trip to the ``Main Mode'' reported in the NetMob trip database~\cite{Chasse2025NetMob25}.
We group these modes into the four categories used throughout the analysis: Vehicle, Bicycle, Walking, and Bus/Public Transportation (Table~\ref{tab:mode-breakdown}).
To represent the GPS observations as paths through the transportation network, we map-match each GPS trace to the OSM road network using a Hidden Markov Model approach~\cite{Newson2009Hidden}, as implemented in the Valhalla open-source routing engine~\cite{2026Valhalla}.
The underlying network is the OSM street network for \idf{} obtained from GeoFabrik and dated January~1, 2022~\cite{OpenStreetMapcontributors2017Planet}.
We retain OSM nodes and ways associated with the ``highway'' tag to construct the surface road network used in the analysis.
For each trip, map matching assigns the GPS observations to a sequence of road-network segments.
We convert this matched sequence into a network path by retaining the corresponding ordered sequence of nodes and removing consecutive duplicate node occurrences.
The matcher failed to produce a valid path for 3{,}692 input trajectories, which were excluded from subsequent analysis.
After restricting the data to the transportation modes considered here, 51{,}318 observed paths remained before network simplification.
The simplification procedure described below reduced this number to 50{,}661 paths used in the primary analyses.

\subsection*{Simulated mobility data}
Simulated mobility is generated using MATSim, an iterative agent-based framework for transportation simulation~\cite{Nagel2009Agentbased,Chakirov2014Enriched,Horni2016MultiAgent,Hackl2019Epidemic}.
We use the open-source Open \idf{} synthetic population and MATSim implementation developed by H\"orl et al.~\cite{Horl2021Synthetic}.
The synthetic population represents individuals with demographic attributes and daily activity plans derived from census, survey, and other public data sources~\cite{Horl2021Synthetic}.
During simulation, agents make transportation decisions including departure-time and route choices, which are iteratively updated within the MATSim framework.
We extract road-based movements from the MATSim event log and group consecutive movement events by agent and trip to obtain paths through the simulation network.
Because MATSim routes agents on a simplified representation of the OSM network, we subsequently project these paths onto the same detailed OSM network used for the observed data.
Nodes retained by MATSim preserve their original OSM identifiers, allowing the two network representations to be aligned directly.
Whenever two consecutive MATSim nodes are not connected by an edge in the detailed comparison network, we interpolate the missing segment using Dijkstra's shortest-path algorithm.
Trips for which no connecting path can be identified are excluded.
MATSim treats walking and bicycle legs as teleportation between their origins and destinations and therefore does not record the route followed through the network.
For these two modes, we reconstruct a network path between the recorded origin and destination using Dijkstra's shortest-path algorithm.
This reconstruction is an analytical assumption rather than an explicit output of the MATSim behavioral model, and its implications are considered in the Discussion.
For public transportation, we retain only bus travel because the analysis is restricted to road-based modes.
The Open \idf{} simulation uses General Transit Feed Specification (GTFS) schedules for public transportation.
We map each bus stop to its associated road-network link and recover the corresponding sequence of OSM nodes to construct bus paths.
We run the Open \idf{} simulation using a population sample fraction of 0.00175, selected to generate a number of simulated paths comparable to the observed dataset.
This procedure yields 54{,}086 simulated paths before network simplification and 53{,}716 paths after simplification.
The final path counts by transportation mode are reported in Table~\ref{tab:mode-breakdown}.

\subsection*{Baseline path construction}
We construct two baseline datasets for each observed and simulated path dataset to provide reference models of network traversal.
Each original path is paired with a memoryless random walk and a shortest path, allowing the baselines to preserve selected properties of the corresponding mobility data.
For the \emph{memoryless random-walk} baseline, we generate an unbiased first-order random walk that begins at the same node and contains the same number of nodes as the corresponding mobility path.
At each step, the next node is selected from the outgoing neighbors of the current node without weighting by edge length or visitation frequency.
For the \emph{shortest-path} baseline, we compute a shortest path through the full OSM comparison network between the origin and destination of each corresponding mobility path.
If several paths have the same minimum cost, one is selected uniformly at random.
The random-walk baseline therefore preserves the origin and initial path-length distribution of the corresponding mobility data but not its destination distribution.
The shortest-path baseline preserves the origin and destination distributions but does not preserve path length.
We construct both baselines before applying the network-simplification procedure described below.
Consequently, simplification and subsequent removal of paths containing fewer than two nodes introduce small differences between the final baseline distributions and those of the original datasets.

\subsection*{Network simplification}
The detailed OSM representation contains many nodes that encode road geometry but do not correspond to meaningful route-choice points.
To focus the analysis on network locations at which movement can branch, we remove such redundant nodes before computing path-level statistics~\cite{Boeing2025Topological}.
For bidirectional road segments, we identify a node as redundant when its in- and out-neighborhoods contain the same two adjacent nodes.
For a one-way road, a node is considered redundant when it has one incoming and one outgoing neighbor and removing it does not alter connectivity between the neighboring nodes.
All redundant nodes are identified on the original network before any nodes are removed so that classification does not depend on the order of simplification.
We then remove the identified nodes from all observed, simulated, and baseline paths.
Paths containing fewer than two nodes after simplification are discarded.
The original comparison network contains 3{,}016{,}582 nodes and 6{,}655{,}074 edges.
After simplification, the network contains 788{,}018 nodes and 2{,}197{,}946 edges, corresponding to 26\% and 33\% of the original node and edge counts, respectively.
Unless otherwise stated, all results reported in the main text are computed using these simplified paths and network.

\subsection*{Path length and subpath diversity}
We represent each mobility path as an ordered sequence of network nodes \(x=(v_1,\ldots,v_n)\).
Path length is characterized both by the number of traversed network nodes and by physical distance, computed from the geometry of the corresponding OSM edges.
Unless stated otherwise, path length and subpath length refer to the number of nodes in the corresponding node sequence rather than physical distance.
For each quantile \(q\), we compare physical-distance distributions using the ratio \(Q_{\mathrm{sim}}(q)/Q_{\mathrm{obs}}(q)\), with analogous ratios computed for the corresponding shortest-path and random-walk baselines (Fig.~\ref{fig:lengths}B).
To examine the relationship between physical distance and network length, we group paths into logarithmically spaced node-length bins and compute the median physical distance per node within each bin (Fig.~\ref{fig:lengths}C).
To quantify sequential diversity, we enumerate all distinct contiguous subpaths of a given length \(L\) across each dataset.
Because the number of subpaths that can be observed depends strongly on the full path-length distribution, we compare the number of distinct subpaths with the total number of contiguous subpath occurrences implied by that distribution.
Let \(\eta_{L'}\) denote the number of complete paths of length \(L'\), and let \(L^{\max}\) denote the maximum observed path length.
The total number of contiguous subpath occurrences of length \(L\) is
\begin{equation}
T_L =
\sum_{L \leq L' \leq L^{\max}}
\eta_{L'}(L'-L+1),
\end{equation}
because a path of length \(L'\) contributes \(L'-L+1\) such occurrences.
If every occurrence corresponded to a distinct subpath, \(T_L\) would also be the maximum possible number of unique subpaths implied by the path-length distribution.
Let \(U_L\) denote the number of distinct subpaths actually observed at length \(L\).
We define the repeat fraction as
\begin{equation}
R_L = 1-\frac{U_L}{T_L},
\end{equation}
where \(R_L=0\) indicates that every subpath occurrence is unique, whereas larger values indicate increasing repetition or overlap (Fig.~\ref{fig:diversity}B).
As an upper bound on the number of unique subpaths, \(T_L\) ignores constraints imposed by the underlying road-network topology and can therefore exceed the number of distinct subpaths that are topologically realizable, particularly at shorter lengths.
To additionally characterize how evenly subpath occurrences are distributed among the distinct sequences observed at each length, let \(n_{i,L}\) denote the occurrence count for subpath \(i\), let \(W_L = \sum_i \eta_{i,L}\) count the total observed visitations over all subpaths of length \(L\), and \(p_{i,L}=\frac{n_{i,L}}{W_L}\) be the relative frequency of subpath \(i\).
We compute the Shannon entropy
\begin{equation}
H_L=-\sum_{i=1}^{U_L}p_{i,L}\ln p_{i,L},
\end{equation}
and define the normalized effective diversity as
\begin{equation}
D_L^{\mathrm{norm}}
=
\frac{\exp(H_L)}{U_L}.
\end{equation}
A value of \(D_L^{\mathrm{norm}}=1\) indicates that all distinct subpaths occur equally often, whereas smaller values indicate increasing concentration of occurrences among a subset of subpaths (Fig.~\ref{fig:diversity}C).

\subsection*{Network coverage and path alignment}
For each dataset, network coverage is measured as the fraction of nodes and edges in the simplified OSM network that appear in at least one path.
We additionally count the total number of occurrences of every node and edge across all paths to characterize differences in visitation frequency.
For each visitation threshold \(m\), we restrict attention to network elements visited at least \(m\) times in one dataset and compute the fraction that is never visited in the other dataset.
For the node-level fractions shown in Fig.~\ref{fig:coverage}B, 95\% confidence intervals are calculated using the Wilson score interval.
To compare the sets of network elements and longer subpaths used by two datasets A and B, we compute Jaccard similarity.
For sets of unique subpaths \(S_L^{A}\) and \(S_L^{B}\) at length \(L\), the Jaccard index is
\begin{equation}
J_L(A,B)
=
\frac{\left|S_L^{A}\cap S_L^{B}\right|}
{\left|S_L^{A}\cup S_L^{B}\right|}.
\end{equation}
The same calculation applied to sets of visited nodes quantifies overlap in the physical network used by the datasets.
We additionally compute a weighted Jaccard index that incorporates the number of times each subpath occurs.
For occurrence counts \(\eta_{i,L}^A\) and \(\eta_{i, L}^{B}\), the weighted similarity is
\begin{equation}
J^{w}_L(A,B)
=
\frac{\sum_i \min\left(\eta_{i,L}^A,\eta_{i, L}^{B}\right)}
{\sum_i \max\left(\eta_{i,L}^A,\eta_{i, L}^{B}\right)},
\end{equation}
where a subpath \(i\) not observed in a dataset  has occurrence count 0.

Finally, we use Kendall's \(\tau\) (specifically Kendall's \(\tau\)\textendash{}\(b\) as it accounts for ties) to quantify whether the relative frequency ranking of subpaths observed at least once in both datasets is preserved as subpath length increases.
Estimates are shown faded in Fig.~\ref{fig:alignment}D where fewer than \(10^4\) subpaths are shared or the correlation is not significant at \(p < 0.05\).

\subsection*{Higher-order network representation and order estimation}
We represent sequential dependencies in mobility paths using higher-order de Bruijn graphs~\cite{Scholtes2017When,Lambiotte2019networks}.
In a model of order \(k\), a higher-order node represents an observed sequence of \(k\) physical network nodes~\cite{Zhang2025Connectivity},
\[
\hat{u}=(v_1,\ldots,v_k),
\]
and a directed transition connects two higher-order nodes when their sequences overlap by \(k-1\) nodes.
The edge weight is the number of times the corresponding sequence of \(k+1\) physical nodes occurs in the path data.
Normalizing outgoing edge weights therefore defines the empirical probability of the next network location conditional on the preceding path history.
For a higher-order state
\(\hat{u}=(v_1,\ldots,v_k)\),
the transition probability to
\(\hat{w}=(v_2,\ldots,v_k,v_{k+1})\)
is
\begin{equation}
P(\hat{w}\mid\hat{u})
=
\frac{W_{\hat{u}\hat{w}}^{(k)}}
{\sum_{\hat{j}\in\mathcal{N}(\hat{u})}
W_{\hat{u}\hat{j}}^{(k)}},
\end{equation}
where \(W_{\hat{u}\hat{w}}^{(k)}\) is the observed transition count and \(\mathcal{N}(\hat{u})\) denotes the outgoing neighbors of \(\hat{u}\).
We estimate the maximum memory order supported by each path dataset using the multi-order model likelihood framework of Scholtes~\cite{Scholtes2017When}.
This approach compares nested models of increasing maximum order while accounting for the corresponding increase in model complexity.
Optimal orders are estimated independently for each transportation mode, the combined path datasets, and their corresponding baselines.

\subsection*{Next-step prediction}
We quantify path predictability using a next-step prediction task based on the empirical transition probabilities of the higher-order network models~\cite{Zhang2025Connectivity}.
Given an observed path prefix, the model predicts which topologically accessible node is most likely to be visited next.
For a path prefix
\((v_1,\ldots,v_\ell)\)
and a model with memory length \(k\), we use the most recent \(k\) nodes to identify the corresponding higher-order state and select the outgoing transition with the largest empirical transition probability.
If that context was not observed in the training data, we successively back off to models of order \(k-1,k-2,\ldots,1\).
If no empirical transition is available at first order, prediction defaults to an unbiased random choice among the outgoing neighbors of the current physical node.
For the primary prediction experiment, we stratify paths according to their order of magnitude in node length and randomly assign 50\% of the paths within each stratum to training and the remaining 50\% to testing.
This stratification preserves the heavy-tailed path-length distribution across the two subsets.
The procedure is repeated independently for ten train--test splits.
For every test path \(x=(v_1,\ldots,v_\ell)\) with \(\ell \geq k+1\), we slide a window of \(k+1\) nodes along the path.
For each window, the first \(k\) nodes are used as the observed history and the final node is treated as the prediction target.
Prediction accuracy is the fraction of these next-step targets predicted correctly across all eligible windows in the test set.
We repeat this analysis for models of increasing maximum memory length.
In a second experiment, we fix the memory length at \(k=2\) and vary the fraction of paths used for training to determine how prediction performance depends on training-set size.
The remaining paths are used for testing at each training fraction, and the same stratification procedure is applied.
At zero training data, prediction reduces to the memoryless fallback rule described above.

\subsection*{Conditional entropy}
Prediction accuracy measures whether the most probable next step is selected correctly but does not capture the full uncertainty of the next-step distribution.
We therefore complement the prediction analysis with the conditional entropy of the empirical higher-order transition probabilities.
For a path context
\(s=(v_1,\ldots,v_k)\),
let
\(\hat{p}(v_{k+1}\mid s)\)
denote the empirical distribution over possible next nodes.
The conditional entropy associated with that context is
\begin{equation}
H(\hat{p}_s)
=
-\sum_{v}
\hat{p}(v\mid s)
\log \hat{p}(v\mid s).
\end{equation}
For this conditional-entropy calculation, we use logarithms to base 2, so entropy is reported in bits.
Let \(n_s\) denote the number of occurrences of context \(s\), and let
\(N_k=\sum_s n_s\)
be the total number of length-\(k\) contexts.
We compute the occurrence-weighted average conditional entropy at memory length \(k\) as
\begin{equation}
H_k
=
\sum_{s\in\mathcal{M}_k}
\frac{n_s}{N_k}H(\hat{p}_s).
\end{equation}
Lower values of \(H_k\) indicate that a given path history is associated with a smaller or more concentrated set of likely next steps, whereas larger values indicate greater uncertainty in the continuation.
We compute \(H_k\) independently for each observed and simulated transportation mode and for the combined datasets across the same range of memory lengths used in the prediction analysis.

\clearpage 

%
\bibliography{references} %
\bibliographystyle{sciencemag}

\section*{Acknowledgements}
We thank the organizers and data providers of the 2025 NetMob Data Challenge for providing access to the dataset analyzed in this work. We also thank Sebastian Horl for the development and helpful discussion of the Open \idf{} MATSim synthetic population and simulation.

\paragraph*{Funding}
This work was supported by a grant from the Fund for Energy Research with Corporate Partners, administered by the Andlinger Center for Energy and the Environment at Princeton University.

\paragraph*{Competing interests}
The authors declare that they have no competing interests.

\paragraph*{Data, code and materials availability}
The NetMob 2025 Data Challenge dataset analyzed in this study is access-controlled; it was made available to challenge participants, cannot be redistributed by the authors, and must be requested from the challenge organizers~\cite{Chasse2025NetMob25}.
The OpenStreetMap road network for the \idf region is openly available via GeoFabrik (\url{https://download.geofabrik.de/europe/france/ile-de-france.html}).
The Open \idf synthetic population and MATSim simulation pipeline is open source and available from H\"orl et al.~\cite{Horl2021Synthetic}.

\paragraph*{Authors' contributions}
T.L.~wrote the code for and carried out the primary data analysis. C.Z.~provided code for the initial data collection, cleaning, and simulation. T.L., C.Z.~and J.H.~designed the research, analyzed and discussed the results, and wrote the paper.


\end{document}